\documentclass[12pt]{article}
\usepackage{graphicx}
\usepackage{amsfonts}
\usepackage{amssymb,amsmath}
\usepackage{latexsym}
\usepackage{color}
\usepackage{cite}
\usepackage{bm}
\input{colordvi.tex}

\renewcommand{\thefootnote}{\#\arabic{footnote}}

\begin{document}

\renewcommand{\thepage}{\arabic{page}}
\setcounter{page}{1}
\renewcommand{\thefootnote}{\#\arabic{footnote}}

\begin{titlepage}

\begin{center}

\vskip .5in

{\Large \bf Precise Measurements of Primordial Power Spectrum  \\ \vspace{1mm}
with 21 cm Fluctuations
}

\vskip .45in

{\large
Kazunori~Kohri$\,^{1,2}$,
Yoshihiko~Oyama$\,^1$, 
Toyokazu~Sekiguchi$\,^3$ \\ and
Tomo~Takahashi$\,^4$
}

\vskip .45in

{\em
$^1$
The Graduate University for Advanced Studies (Sokendai), 1-1 Oho, Tsukuba 305-0801, Japan\\
$^2$
Institute of Particle and Nuclear Studies, KEK, 1-1 Oho, Tsukuba 305-0801, Japan\\
$^3$
Graduate School of Science, Nagoya University,
Furo-cho, Chikusa-ku, Nagoya,  464-8602, Japan\\
$^4$
Department of Physics, Saga University, Saga 840-8502, Japan
}

\end{center}

\vskip .4in

\begin{abstract}

  We discuss the issue of how precisely we can measure the primordial
  power spectrum by using future
  observations of 21 cm fluctuations and cosmic microwave background
  (CMB).   For this purpose, we investigate projected constraints on the quantities
  characterizing primordial power spectrum: the spectral index $n_s$,
  its running $\alpha_s$ and even its higher order running
  $\beta_s$. We show that future 21 cm observations in combinations
  with CMB would accurately measure above mentioned observables of
  primordial power spectrum. We also discuss its implications to some explicit
  inflationary models. 
  
\end{abstract}
\end{titlepage}

\setcounter{footnote}{0}

\section{Introduction}

It is now widely believed that the Universe experienced an extremely
rapid expanding phase, the inflation, in the early stage of its
history, which is assumed to be driven by the vacuum energy of a
scalar field, called inflaton. In addition that the inflaton gives an
inflationary expanding era, it can also provide the source of
primordial density fluctuations, whose nature can be utilized to
differentiate inflationary models.  Specifically, the spectral index
$n_s$, describing the scale dependence of the primordial power
spectrum, and the tensor-to-scalar ratio $r$, characterizing the
amplitude of the gravity waves, are usually discussed, which can be
measured/constrained by cosmological observations such as cosmic
microwave background (CMB), large scale structure and so on. In
particular, in the near future, several experiments are expected to
measure/constrain the tensor-to-scalar ratio with high sensitivity,
which may enable us to grasp invaluable hints to understand the
inflationary model realized in the nature.

However, on the other hand, there exist inflation models which predict
too small tensor-to-scalar ratio to be detected, at least, in near
future experiments. Furthermore, primordial density fluctuations are
not necessarily generated from the inflaton, but they can be produced
by 
a light scalar field other than the inflaton, such as in the
curvaton model \cite{Enqvist:2001zp,Lyth:2001nq,Moroi:2001ct},
modulated reheating scenario \cite{Dvali:2003em,Kofman:2003nx} and so
on. In particular, models of these kind have been attracting attention
since they can predict large values of $f_{\rm NL}$, a measure of
bispectrum which characterizes non-Gaussian nature of density
fluctuations\footnote{
  Although the standard inflation models generally predict $f_{\rm NL}
  \ll \mathcal{O}(1)$, current observations of CMB and large scale
  structure give the constraints on $f_{\rm NL}$ for the local type as $ -3 < f_{\rm NL}
  < 77 ~{\rm (95 \% ~C.L.)}$ from WMAP9 \cite{Bennett:2012fp} and $ 25
  < f_{\rm NL} < 117 ~{\rm (95 \% ~C.L.)}$ from NRAO VLA Sky Survey
  \cite{Xia:2010yu}.
}. In these models, the tensor-to-scalar ratio is generally very
small, and thus one may not be able to detect any
gravity wave signal if such a model is the correct one for generating
density fluctuations.  In such a case, we may need to pursue another
quantity which is not related to the gravity waves.  As such a
quantity, the spectral index $n_s$ is widely
discussed.  However, there are many inflationary
models which can predict almost the same value of $n_s$, thus we need
something more.

In light of these considerations, it would be desirable to have more
observables, in particular, which are irrelevant to the tensor
mode. Needless to say, such a quantity would also be helpful even if
we can have information from the gravity waves as well, and give
complementary information.  As such an observable, we consider the
running of the spectral index, $\alpha_s$ and even a higher order
running or ``the running of the running" $\beta_s$.  The former
quantity, $\alpha_s$, has already been explored by many authors.
However, the latter one, $\beta_s$, has not been
investigated much in the literature\footnote{
See Refs. \cite{Powell:2012xz,Huang:2006hr} for examples of such works.
}, on which we especially focus in this paper.  In general, higher
order runnings are expected to be very small compared to the spectral
index $n_s - 1$, thus it seems very difficult to actually measure such a
quantity.  However, as we will show in this paper, future
experiments of 21 cm fluctuation can give precise measurements of
$\alpha_s$ and $\beta_s$, which would be very useful to differentiate
inflationary models.  In fact, the issue of probing the primordial
power spectrum with 21 cm experiments was discussed in Refs.
\cite{Mao:2008ug,Barger:2008ii}. In this paper, we further extend the
discussion including the running of the running $\beta_s$ and study
its expected constraints from future observations of 21 cm
fluctuations in combination with CMB. In addition to observational
constraints, we also study the runnings $\alpha_s$ and $\beta_s$ as
well as $n_s$ in some explicit models and discuss its testability and
in what cases the information from $\alpha_s$ and $\beta_s$ will be
useful to differentiate inflationary models.

This paper is organized into four sections as follows. In the next section, we give a formalism to discuss the
power spectrum of primordial density fluctuations, paying particular
attention to its scale dependence such as $n_s, \alpha_s$ and
$\beta_s$. We also give the predictions for these variables in some
explicit models and discuss in what models/cases the information of
the runnings are useful.  In Section~\ref{sec:obs}, we describe our
method to derive expected constraints from future observations of 21
cm fluctuations and CMB and present out results.  The final section is
devoted to summary of this paper.

\section{Inflationary parameters: Formalism}

As mentioned in the introduction, the purpose of this paper is to
investigate what kind of information can be obtained by considering 
the scale dependence of primordial curvature
perturbations including  ``higher order"  ones.
Before going to study expected constraints, here in this
section, we give a formalism and formulas to discuss primordial
curvature perturbations and its scale dependence. We also study some
explicit inflationary models.  
Since primordial curvature perturbation can also be generated from 
a light scalar field other than the inflaton such as in the curvaton, modulated 
reheating scenarios and so on, we give the formulas for such cases as well.
But first we start
with a case of the standard inflation model.

\subsection{Standard inflation case}

For the standard inflation case, the primordial curvature power spectrum is given, in terms of the $\delta N$ formalism, by
\begin{equation}
\label{eq:P_zeta_inf}
{\cal P}_\zeta  (k)
= 
\left( \frac{\partial N}{\partial \phi_\ast} \right)^2 \mathcal{P}_{\delta \phi} (k),
\end{equation}
where $\mathcal{P}_{\delta \phi} = (H_\ast / 2\pi)^2 \left( 1+ 2 (3C - 1) \epsilon - 2 C\eta \right)$
is the power spectrum for fluctuations of the inflaton field $\phi$  \cite{Stewart:1993bc,Byrnes:2006fr}. 
For the standard inflation case, $\partial N / \partial \phi_\ast = - H_\ast / \dot{\phi}_\ast$ 
with a dot representing
derivative with respect to time.  The subscript ``$\ast$" indicates
that the quantity is evaluated at the time of horizon exit $k=a_\ast H_\ast$ and
$C = 2 - \ln 2 -  b \simeq 0.73$ with $b$ being the Euler-Mascheroni
constant.  $\epsilon$ and $\eta$ are slow-roll parameters which are
defined as
\begin{eqnarray}
\epsilon & = &  \frac12 M_{\rm pl}^2 \left( \frac{V_\phi}{V} \right)^2,\\ \notag \\
\eta  & = &  M_{\rm pl}^2 \frac{V_{\phi\phi}}{V},
\end{eqnarray}
where $V$ is the potential for the inflaton and a subscript $\phi$
represents a derivative with respect to the inflaton field.  For later
use, here we also define higher order slow-roll parameters:
\begin{eqnarray}
\xi^{(2)}  & = &  M_{\rm pl}^4 \frac{V_\phi V_{\phi\phi\phi}}{V^2}, \\ \notag \\ 
\sigma^{(3)}  & = &  M_{\rm pl}^6 \frac{(V_\phi)^2 V_{\phi\phi\phi\phi}}{V^3},
\end{eqnarray}
where we have put the superscripts $(2)$ and $(3)$ to remind that
$\xi^{(2)}$ and $\sigma^{(3)}$ are second and third order slow-roll
quantities, respectively.

We expand the power spectrum in terms of the logarithm of the wave
number up to the 3rd order as
\begin{eqnarray}
{\cal P}_{\zeta} (k) &=& {\cal P}_{\zeta} (k_{\rm ref}) 
\exp 
\left[ 
(n_s -1 ) \ln  \left( \frac{k}{k_{\rm ref}} \right)
+ \frac12 \alpha_s \ln^2  \left( \frac{k}{k_{\rm ref}}  \right)
+ \frac{1}{3 !}  \beta_s \ln^3 \left( \frac{k}{k_{\rm ref}}  \right)
\right] \notag \\ \notag \\
&=& 
 {\cal P}_{\zeta} (k_{\rm ref}) 
 \left( \frac{k}{k_{\rm ref}}  \right)^{
 n_s -1 + \frac12 \alpha_s  \ln (k /k_{\rm ref})  + \frac16 \beta_s  \ln^2 (k /k_{\rm ref}) 
 },
\end{eqnarray}
where we have defined 
\begin{eqnarray}
\alpha_s & = &  \frac{d n_s}{d \ln k}, \\
\notag \\
\beta_s & = & \frac{d \alpha_s}{d\ln k}.
\end{eqnarray}
$\alpha_s$ is usually called ``running" of $n_s$. Thus in this
sense, $\beta_s$ may be called as ``the running of the running."

By using the slow-roll parameters, $n_s, \alpha_s$ and $\beta_s$ are
written as 
\begin{eqnarray}
n_s -1  
& = &  
-6 \epsilon + 2 \eta 
+ \left( - \frac{10}{3} + 24 C \right) \epsilon^2 + \frac23 \eta^2 - (2+ 16C) \epsilon \eta + \left( \frac23 + 2C \right) \xi^{(2)}, \\
\alpha_s & = &   -24 \epsilon^2 + 16 \epsilon \eta - 2 \xi^{(2)},  \\ \notag \\
\beta_s &=& 
-192 \epsilon^3 + 192 \epsilon^2 \eta  -32 \epsilon \eta^2 
+ ( -24 \epsilon + 2 \eta ) \xi^{(2)}  + 2 \sigma^{(3)}.
\end{eqnarray}

\subsection{Case with a light scalar field}

When the primordial curvature perturbation originates to fluctuations
of a light scalar field $\sigma$ other than the inflaton, such as in
the case of the curvaton and modulated reheating models, the expressions for the spectral index
and its runnings are different from those in the standard inflation
case. Here we assume that the energy density of such a light scalar field
is subdominant during inflation, and hence it does not affect the
inflationary dynamics.  To make our discussion general, we consider
the case where the inflaton and a light scalar field can be both
responsible for density perturbations\footnote{
  Models of this kind are called ``mixed" model. Such models have been
  investigated in the context of the curvaton model
  \cite{Langlois:2004nn,Lazarides:2004we,Moroi:2005kz,Moroi:2005np,Ichikawa:2008iq}
  and modulated reheating scenario \cite{Ichikawa:2008ne}.
}. Cases with the standard inflation or a light scalar field being purely
responsible for density perturbations are given as limiting cases.

Denoting a light field as $\sigma$ and assuming that the inflaton and
the $\sigma$ field are uncorrelated, the total curvature perturbation
can be written as
\begin{equation}
\mathcal{P}_\zeta  (k) = \mathcal{P}_\zeta^{(\phi)}(k) + \mathcal{P}_\zeta^{(\sigma)}(k),
\end{equation}
where $\mathcal{P}_\zeta^{(\phi)}$ and $\mathcal{P}_\zeta^{(\sigma)}$ respectively
represent the contribution from the inflaton and $\sigma$ fields.
The inflaton part $ \mathcal{P}_\zeta^{(\phi)}$ is actually the one given in 
Eq.~\eqref{eq:P_zeta_inf}.  For the contribution from a light scalar
field, $\mathcal{P}_\zeta^{(\sigma)}$ is given by 
\begin{equation}
{\cal P}_\zeta^{(\sigma)}  (k) 
= 
\left( \frac{\partial N}{\partial \sigma_\ast} \right)^2 \mathcal{P}_{\delta \sigma}(k),
\end{equation}
where 
$\mathcal{P}_{\delta \sigma} = (H_\ast / 2\pi)^2 \left(  1- 2 (1 - C) \epsilon - 2 C\eta_\sigma \right)$
is the power spectrum for fluctuations of a light scalar field $\sigma$ \cite{Byrnes:2006fr}. Here
the slow-roll parameter $\eta_\sigma$ for $\sigma$ field is defined
similarly to those for the inflaton case as
\begin{equation}
\eta_\sigma   =    \frac{U_{\sigma\sigma}}{3 H_\ast^2}, 
\end{equation}
where we assume that the scalar potential is given by the
sum of those for $\phi$ and $\sigma$ as $V(\phi) + U(\sigma)$.   For
later use, here we also define other higher order slow-roll parameters
as
\begin{eqnarray}
\xi_\sigma^{(2)}  & = &  \frac{U_\sigma U_{\sigma\sigma\sigma}}{(3 H_\ast^2)^2}, \\ \notag \\ 
\sigma_\sigma^{(3)}  & = &   \frac{(U_\sigma)^2 U_{\sigma\sigma\sigma}}{(3 H_\ast^2)^3},
\end{eqnarray}
where a subscript $\sigma$ denotes the derivative with respect to $\sigma$ field. 

The spectral index $n_s$ and its runnings $ \alpha_s$ and $\beta_s$ in
this case are given as follows: 
\begin{eqnarray}
\label{eq:ns_mix}
n_s -1 &&=
  \Xi_\phi 
\left[ 
-6 \epsilon + 2 \eta 
+ \left( - \frac{10}{3} + 24 C \right) \epsilon^2 + \frac23 \eta^2 - (2+ 16C) \epsilon \eta + \left( \frac23 + 2C \right) \xi^{(2)}
\right] \notag  \\  
&&
+ \Xi_\sigma
\left[
 -2 \epsilon + 2 \eta_\sigma 
 + \left( - \frac{22}{3} + 8C \right)  \epsilon^2 + \frac23 \eta_{\sigma}^2 
  + \left( \frac83 - 4C \right) \epsilon \eta  +  \left( \frac23 - 4C \right) \epsilon \eta_{\sigma}  
  \right. \notag \\
  && 
  \left.
 + \frac43 \xi^{(2)}  + 2C \xi_\sigma^{(2)}
 \right], 
\end{eqnarray}
\begin{eqnarray}
\alpha_s &&=  \Xi_\phi (-24 \epsilon^2 + 16 \epsilon \eta - 2 \xi_\phi^{(2)} ) 
+ \Xi_\sigma (- 8\epsilon^2 + 4 \epsilon \eta_\phi + 4 \epsilon \eta_\sigma - 2 \xi_\sigma^{(2)} ) \notag\\
&& - 4 \Xi_\phi \Xi_\sigma ( 4 \epsilon + 2 \eta_\phi - 2 \eta_\sigma)^2,  \\ \notag \\
\beta_s   &&  =
\Xi_\phi (-192 \epsilon^3 + 192 \epsilon^2 \eta  -32 \epsilon \eta^2 
+ ( -24 \epsilon + 2 \eta ) \xi^{(2)}  + 2 \sigma^{(3)} ) \notag \\
&& \Xi_\sigma( -64 \epsilon^3 + 56 \epsilon^2 \eta_\phi - 8 \epsilon \eta_\phi^2 
+ 24 \epsilon^2 \eta_\sigma - 8 \epsilon \eta_\phi \eta_\sigma  -12 \epsilon \xi_\phi^{(2)}
- 4 \epsilon \xi_\sigma^{(2)} + 2 \eta_\sigma \xi_\sigma^{(2)} + 2 \sigma^{(3)}_\sigma) ,\notag \\\notag
&&
12 \Xi_\phi \Xi_\sigma (2 \epsilon - \eta + \eta_\sigma) ( 8 \epsilon^2 - 6 \epsilon \eta_\phi + 2 \epsilon \eta_\sigma + \xi^{(2)} - \xi_\sigma^{(2)} )
\notag \\
&&
+8 \Xi_\phi \Xi_\sigma  ( \Xi_\phi - \Xi_\sigma) (2\epsilon - \eta + \eta_\sigma )^3,
\end{eqnarray}
where we have defined the fraction of $\mathcal{P}_\zeta^{(\phi)}$ and
$\mathcal{P}_\zeta^{(\sigma)}$ in the total power spectrum, respectively, as
\begin{equation}
\Xi_\phi \equiv \frac{ \mathcal{P}_\zeta^{(\phi)}}{ \mathcal{P}_\zeta^{(\phi)} + \mathcal{P}_\zeta^{(\sigma)} },
\qquad \qquad
\Xi_\sigma \equiv \frac{ \mathcal{P}_\zeta^{(\sigma)}}{\mathcal{P}_\zeta^{(\phi)} + \mathcal{P}_\zeta^{(\sigma)} }. 
\end{equation}
By definition, the sum of these quantities is unity, $\Xi_\phi + \Xi_\sigma = 1$.

\subsection{Some example}

Now we discuss some explicit models and their predictions for the
spectral index $n_s$ and the runnings $\alpha_s$ and $\beta_s$.  To
provide an example of in what cases the information from the runnings
is useful, here we consider models with small tensor-to-scalar
ratio, in other words, the information from the (scalar-mode) power spectrum only available.
We will argue below that, in some cases, 
 ``running of the running" can be very useful to discriminate inflationary models. 
For descriptive purpose, we consider three models and discuss their predictions on $n_s,
\alpha_s$ and $\beta_s$.

\bigskip
\bigskip
\noindent 
\begin{description}
\item[Model I] {\bf : Curvaton with quartic chaotic inflation model }

  In this model, we assume that the curvaton field is totally
  responsible for the cosmic density perturbations. However, even in
  the curvaton model, the inflationary expansion is driven by the
  inflaton field and we need to specify the inflation model, or the
  inflaton potential, to compute the power spectrum. For the inflation
  model, we adopt the chaotic inflation model with a quartic
  potential:
\begin{equation}
V(\phi) = \lambda \phi^4,
\end{equation}
where $\lambda$ is a coupling parameter.  In fact, the quartic chaotic
inflation is already excluded by current cosmological observations
\cite{Hinshaw:2012fq} when  the inflaton is completely responsible for
density perturbations. However, when the curvaton generates primordial
density fluctuations, it is still viable, and even more worth
mentioning, such an inflation model is preferable for this case
 \cite{Moroi:2005kz,Ichikawa:2008iq} since it can give the
spectral index of $n_s \sim 0.96$,  which is close to the central value
from observational constraints.  We should also note here that, although
the curvaton model has been attracting attention because it can
generate large non-Gaussianity, but it is not necessarily large: the
curvaton can also predict small values of $f_{\rm NL}$ in some
parameter range. Thus, non-Gaussianity may not give
useful information in such a case even for this type of model.

For the curvaton sector, we assume that its mass is very small so that
we can set $\eta_\sigma = 0$ in the calculation of $n_s$ and so on.
Furthermore, as mentioned above, the curvaton is assumed to be
totally responsible for the primordial power spectrum,
and hence we also set $\Xi_\sigma = 1 ~(\Xi_\phi
=0)$.  To realize this situation, fluctuations from the inflaton
should satisfy $P_\zeta^{(\phi)} \ll 2.4 \times 10^{-9}$, which means
that the inflation scale is low and the tensor-to-scalar ratio becomes
very small.

In Fig.~\ref{fig:ns_alpha_beta}, we plot the values of  $n_s, \alpha_s$ and $\beta_s$ 
in this model as well as those for Model II which will be discussed in the 
following. In the figure,  the predictions are shown 
in  the $n_s$--$\alpha_s$ (left panel) and the $\alpha_s$--$\beta_s$ (right panel) planes.
Here the number of $e$-folds at the horizon exit is varied as $40 \le N_e \le 60$.

\bigskip
\bigskip
\item[Model II] {\bf :Mutated hybrid inflation model} 

  For this model, we assume that the inflaton is fully responsible for
  density perturbations and the inflaton potential has the following
  form:
\begin{equation}
V (\phi ) 
= 
V_0 \left[ 1 - \left( \frac{\mu}{\phi} \right)^q + \cdots \right],
\end{equation}
which is called mutated hybrid inflation model \cite{Lyth:1996kt} and $q$ is assumed
to be a positive integer.  $\mu$ is some energy scale which
characterizes the model.

\begin{figure}[tbp]
  \begin{center}
    \resizebox{150mm}{!}{
    \includegraphics{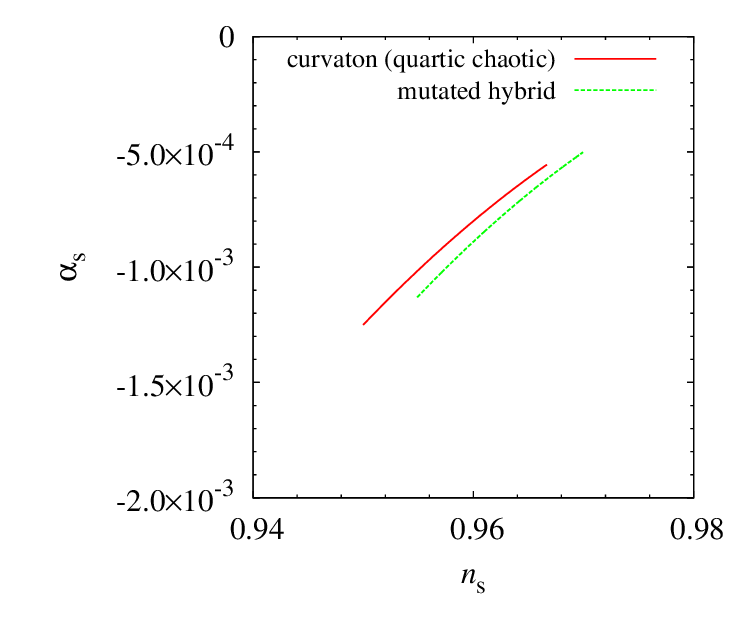}\hspace{10mm}
     \includegraphics{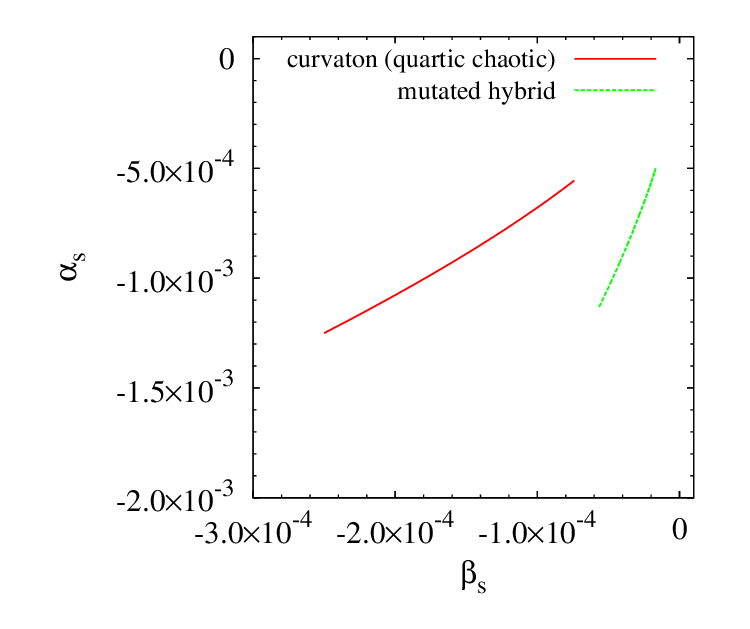}
}
  \end{center}
  \caption{Predictions in the $n_s$--$\alpha_s$ (left) and the
    $\alpha_s$--$\beta_s$ (right) planes from curvaton model with
    chaotic inflation ($V(\phi) = \lambda \phi^4$) and mutated hybrid
    model with $q=4$ and $\mu =  M_{\rm pl}$. The number of
    $e$-folds is varied as $ 40 \le N_e \le 60$ in this figure. 
    }
  \label{fig:ns_alpha_beta}
\end{figure}

\begin{figure}[tbp]
  \begin{center}
    \resizebox{150mm}{!}{
    \includegraphics{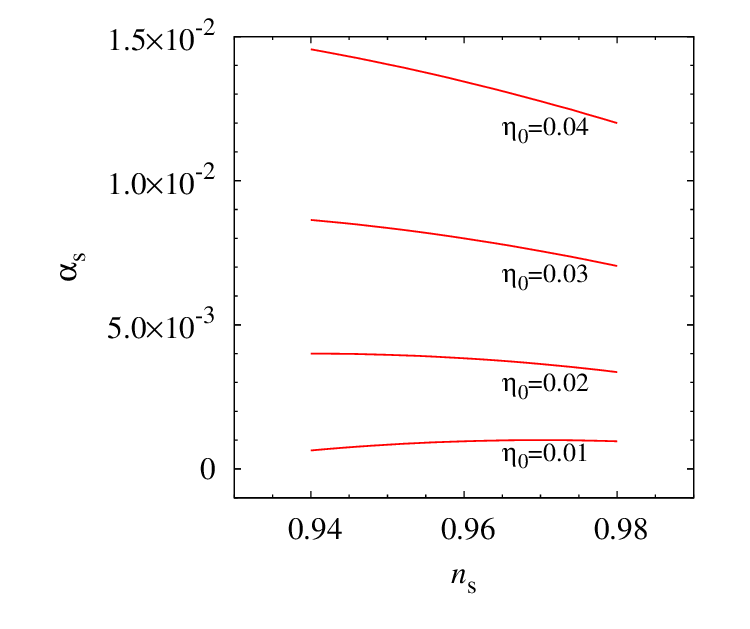}\hspace{10mm}
     \includegraphics{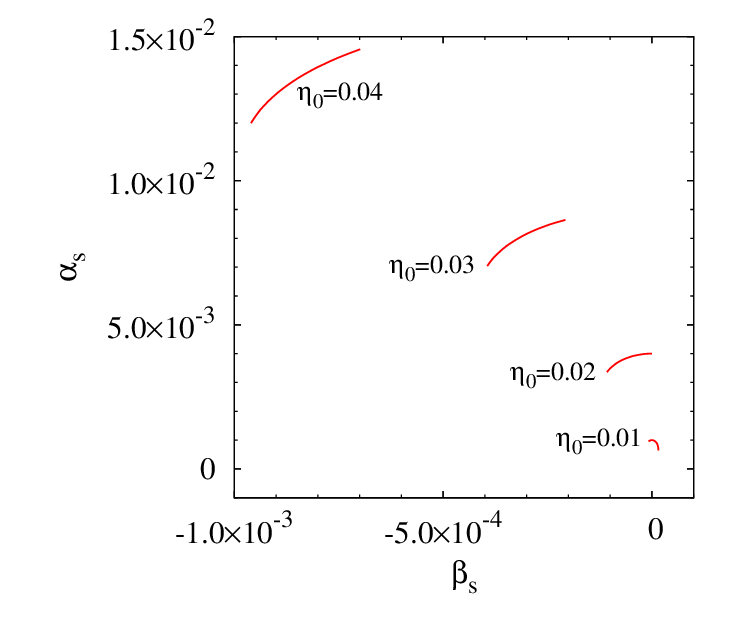}
}
  \end{center}
  \caption{Predictions in the $n_s$--$\alpha_s$ (left) and the
  $\beta_s$--$\alpha_s$ (right) planes from Type III hilltop inflation
  models. From the top to bottom, $\eta_{0}$ is taken to be $0.04$, $0.03$,
  $0.02$, and $0.01$.}
  \label{fig:ns_alpha_beta2}
\end{figure}

In Fig.~\ref{fig:ns_alpha_beta}, we show the predictions for $n_s, \alpha_s$ and $\beta_s$
in this model along with those for Model I.
For  illustration purposes, here we take $q=8$ and
$\mu = M_{\rm pl}$.  In the figure,  $N_e$  is
varied in the range $N_e = 40 -60$ to plot the lines.  We note that,
in both models, the tensor-to-scalar ratio is small as $r < 10^{-3}$.
By looking at the prediction in the $n_s$--$\alpha_s$ plane, one can see that these
models give very similar predictions for $n_s$ and
$\alpha_s$. However, if we look at the $\alpha_s$--$\beta_s$ plane,
their values seems separated compared to those in $n_s$--$\alpha_s$
plane, which shows an idea that a higher order running $\beta_s$ can
give useful information to differentiate inflationary models.  To
state this more quantitatively, we need to know attainable constraints
on these observables from future cosmological observations, which is
the topic in the next section.

\item[Model III]{\bf : Type-III hilltop inflation models}
  
  As a last example,  we discuss a model in which  both $|\alpha_s|$ and
  $|\beta_s|$ can be large. In so-called Type-III hilltop inflation
  models~\cite{German:2001tz, Kohri:2007gq} (see also
  Refs.~\cite{TypeIII-Hilltops,RunnningMass,OtherTypeIIIHilltop} and
  references therein), we expect a large running and a large running
  of running towards small scales. The potential shape is represented
  by
\begin{equation}
    V(\phi) = V_{0} 
            \left( 1 + \frac12 \eta_{0} \frac{\phi^{2}}{M_{\rm pl}^{2}} \right)
            -\lambda \frac{\phi^{q}}{M_{\rm pl}^{q-4}}.
\label{eq:HilltopPotential}
\end{equation}
The field $\phi$ rolls from the top of the hill towards the origin
during the inflationary epoch. Intermediately, the inflation ends by a
water-fall mechanism.  In Fig.~\ref{fig:ns_alpha_beta2}, we plot the
predictions in the $n_s$--$\alpha_s$ (left) and the
$\beta_s$--$\alpha_s$ (right) planes, in case of $q =6$ and
$V_{0}^{1/4}/M_{\rm pl} = 1.6 \times 10^{-7}$. Because the tensor-to-scalar 
ratio is tiny with an order of $r \sim 2 \times 10^{-20}$ in
this model, it would be difficult to detect it. However, with future
precise measurements of $\alpha_{s}$ and $\beta_{s}$, we will be able
to discriminate this model from others.

\end{description}

\newpage

\section{Analysis}
\label{sec:obs}

Now we discuss expected observational constraints on the quantities
such as the spectral index $n_s$, its runnings $\alpha_s$ and 
$\beta_s$ from future cosmological observations by
using Fisher matrix formalism. For this purpose, we investigate a
joint constraint from 21cm fluctuations and CMB. We first briefly
describe the formalism in each below, then present our results on the
future constraints.

\subsection{21cm power spectrum}

Here we briefly sketch and summarize how we can constrain primordial
power spectrum by using 21 cm observations. For detailed descriptions,
we refer the readers to, e.g., Refs. \cite{McQuinn:2005hk,Furlanetto:2006jb,Mao:2008ug, Pritchard:2011xb,Oyama:2012tq}.

The 21 cm line from the hyperfine transition of neutral hydrogen can
be observed as the differential brightness temperature relative to the
CMB temperature $T_{\rm cmb}$:
\begin{equation}
\label{eq:Tb}
T_b ({\bm x})  
= 
\frac
{3 c^3 h A_{10} n_H ({\bm x}) \left( T_s ({\bm x}) - T_{\rm cmb} \right)}
{32 \pi k_B \nu_{21}^2 T_s ({\bm x}) (1+z)^2 (dv_{\parallel} / dr)},
\end{equation}
where $A_{10} \simeq 2.85 \times 10^{-15}{\rm s^{-1}}$ is the spontaneous
decay rate of 21 cm transition and $\nu_{21} $ is the 21 cm line
(rest) frequency.  $T_s({\bm x})$ is the spin temperature, which is
defined by $n_1/ n_0 = 3 \exp ( - T_{21} / T_S)$ with $n_0$ and $n_1$
being respectively the number densities of singlet $1 _0 S_{1/2}$ and
triplet $1 _1 S_{1/2}$ states of neutral hydrogen atom. Here $T_{21}= hc / k_B
\lambda_{21}$ is the temperature corresponding to 21 cm line with
$\lambda_{21}$ being its wavelength. 
$dv_{\parallel} / dr$ is the gradient of the velocity along the line
of sight.

Now we are going to consider fluctuations of $T_b ({\bm x})$.  By
expanding the hydrogen number density $n_H$ and the ionization
fraction $x_i$ (we also sometimes use $x_H = 1 - x_i$: the fraction of
neutral hydrogen) as $n_H({\bm x}) = \bar{n}_H (1 + \delta ({\bm x})
)$ and $x_i({\bm x}) = \bar{x}_i (1 + \delta_x ({\bm x}) )$, we can
rewrite Eq.~\eqref{eq:Tb} as
\begin{equation}
T_b ({\bm x})  
=  
\bar{T}_b  \left( 1 - \bar{x}_i ( 1+ \delta_x ({\bm x}) \right) (1+ \delta ({\bm x}) )
\left( 1 - \frac{1}{Ha} \frac{d v_r}{dr} \right),
\end{equation}
where we have assumed that $T_s \gg T_{\rm cmb}$ since we focus on the
epoch of reionization during which this condition is well satisfied.
$\bar{T}_b$ is the spatially averaged brightness temperature at
redshift $z$ and given by
\begin{equation}
\bar{T}_b \simeq 27~{\rm mK} \left( \frac{\Omega_b h^2}{0.023} \right) \left( \frac{0.15}{\Omega_m h^2} \frac{1+z}{10} \right)^{1/2},
\end{equation}
and $d v_r /dr$ is the peculiar velocity along the line of sight.

By denoting the fluctuation in $T_b$ as 
$\delta T_b ({\bm x}) = T_b({\bm x}) - \bar{x}_{H}\bar{T}_b$,
the 21 cm power
spectrum $P_{21} ({\bm k})$ in the $k$-space is defined by
\begin{equation}
\label{eq:power}
\left\langle  
\delta T^\ast_b ({\bm k}) \delta T_b ({\bm k}')  \right\rangle
= (2\pi)^3 \delta^3 ( \bm{k-k'}) P_{21} ({\bm k}).
\end{equation}
By treating the peculiar velocity $ \delta_v \equiv (dv_r / dr)(1/aH)$
as a perturbation and using that its
Fourier transform can be given by $ \delta_v  ({\bm k})= -\mu^2 \delta  ({\bm k})$ with $\mu = \hat{\bm k}\cdot \hat{\bm n}$
being the cosine of the angle between the wave vector and the line of
sight, the power spectrum can be written as
\begin{equation}
P_{21} ({\bm k}) 
=
P_{\mu^0} (k) + \mu^2 P_{\mu^2} (k)  + \mu^4 P_{\mu^4} (k), 
\end{equation}
where  $k = |{\bm k}|$ and
\begin{eqnarray}
P_{\mu^0}  & = & \mathcal{P}_{\delta \delta} - 2 \mathcal{P}_{x\delta} + \mathcal{P}_{xx},
 \\
P_{\mu^2}   & = & 2 \left( \mathcal{P}_{\delta \delta} - \mathcal{P}_{x\delta} \right),  
\\
P_{\mu^4}  & = &   \mathcal{P}_{\delta \delta}. 
\end{eqnarray}
Here 
$ \mathcal{P}_{\delta \delta} = \bar{T}_b^2 \bar x_H^2P_{\delta \delta},
\mathcal{P}_{x \delta} = \bar{T}_b^2 \bar{x}_i \bar{x}_H P_{x
  \delta} $ and $ \mathcal{P}_{x x} = \bar{T}_b^2 \bar{x}_i^2 P_{ x x} $ 
and $P_{\delta\delta}, P_{x\delta}$ and
$P_{xx}$ are the power spectra defined in the same manner as
Eq.~\eqref{eq:power} for $\delta$ and $\delta_x$.  Since $\delta$
represents fluctuations in the hydrogen number density,
$P_{\delta\delta}$ traces that of matter which includes the
information on primordial power spectrum.  $P_{x\delta}$ and $P_{xx}$
can be neglected if we consider the era when the intergalactic medium
(IGM) is completely neutral. However, after the reionization starts,
in which we are interested, these two spectra contribute
significantly.  Although a rigorous evaluation of these power spectra
may need some numerical simulations, here we adopt the treatment given
in Ref. \cite{Mao:2008ug}, where $\mathcal{P}_{x\delta}$ and
$\mathcal{P}_{xx}$ are assumed to have a specific form to match
radiative transfer simulations of Refs. 
\cite{McQuinn:2006et,McQuinn:2007dy}.  Their explicit forms are the
followings:
\begin{eqnarray}
\label{eq:Pxx}
\mathcal{P}_{x x}  (k) 
& = & 
b_{xx}^2 \left[ 1 + \alpha_{xx} (k R_{xx}) + (k R_{xx})^2 \right]^{-\gamma_{xx} / 2} \mathcal{P}_{\delta\delta} (k), \\
\label{eq:Pxdelta}
\mathcal{P}_{x \delta} (k) 
& = &  
b_{x\delta}^2 ~e^{ - \alpha_{x\delta} (k R_{x\delta}) - (k R_{x\delta})^2} \mathcal{P}_{\delta\delta} (k),
\end{eqnarray}
where $b_{xx}$, $b_{x\delta}$, $\alpha_{xx}$, $\gamma_{xx}$ and $\alpha_{x\delta}$
are parameters which characterize the amplitudes and the shapes of the spectra.
$R_{xx}$ and $R_{x\delta}$ correspond to the
effective size of the ionized bubbles.  The values of these parameters
we adopt in the analysis are listed in Table~\ref{tab:Pxx_xdelta}.

\begin{table}[t]
  \centering 
  \begin{tabular}{ccccccccc}
\hline \hline
~~$z$~~ & ~~$\bar{x}_H$~~
& ~~$b_{xx}^2$~~ & ~~$R_{xx}$~~  & ~~$\alpha_{xx}$~~ & ~~$\gamma_{xx}$~~ 
& ~~$b_{x\delta}^2$~~ & ~~$R_{x\delta}$~~  & ~~$\alpha_{x\delta}$~~ \\ 
&  &  & $[{\rm Mpc}]$& & & &$[{\rm Mpc}]$ \\
\hline
$9.2$  &  $0.9$ & $0.208$ & $1.24$ & $-1.63$ & $0.38$ & $0.45$ & $0.56$   & $-0.4$ \\ 
$8.0$  &  $0.7$ & $2.12$   & $1.63$ & $-0.1$   & $1.35$ & $1.47$ & $0.62$   & $0.46$ \\ 
$7.5$  &  $0.5$ & $9.9$     & $1.3$   & $1.6$    & $2.3$   & $3.1$   & $0.58$   & $2.0$ \\ 
$7.0$  &  $0.3$ & $77.0$   & $3.0$   & $4.5$    & $2.05$ & $8.2$   & $0.143$ & $28.0$ \\
\hline \hline
\end{tabular}
\caption{Fiducial values for the parameters in $\mathcal{P}_{xx}(k)$
  and $\mathcal{P}_{x\delta}(k)$ (See Eqs.~\eqref{eq:Pxx} and
  \eqref{eq:Pxdelta}) \cite{Mao:2008ug}.}
  \label{tab:Pxx_xdelta}
\end{table}

Now we are in the position to discuss the Fisher matrix for 21 cm
observations.  First of all, we note that experiments of 21 cm
radiation do not directly measure the wave number ${\bm k}$ nor the
power spectrum in $k$-space $P_{21} ({\bm k})$.  Instead, an
experiment measures the angular location on the sky and the frequency
which can be specified by the vector
\begin{equation}
{\bm \Theta} = \theta_x \hat{e}_x + \theta_y \hat{e}_y + \Delta f \hat{e}_z (\equiv  {\bm \Theta}_\perp + \Delta f \hat{e}_z ).
\end{equation}
Here the frequency is represented by its difference from the central
redshift $z$ of a given redshift bin.  Then we can define the Fourier
dual of ${\bm \Theta}$ as
\begin{equation}
{\bm u} \equiv u_x \hat{e}_x + u_y \hat{e}_y + u_\parallel \hat{e}_z \left( \equiv {\bm u}_\perp + u_\parallel \hat{e}_z \right).
\end{equation}
Notice that, since $u_\parallel$ is the Fourier dual of $\Delta f$, it
has the units of time.  Assuming that the sky is flat\footnote{
  Even if we consider all-sky experiments, the flat-sky approximation
  can be valid as long as the data are analyzed in many small patches
  of the sky \cite{Mao:2008ug}.  
}, we can linearize the relation ${\bm r}$ and ${\bm \Theta}$.
Denoting the vector perpendicular to the line of sight as ${\bm
  r}_\perp$, we have the relations
\begin{equation}
{\bm \Theta}_\perp = {\bm r}_\perp / d_A (z_\ast), 
\qquad
\Delta f = \Delta r_\parallel / y(z_\ast)
\end{equation}
where $d_A (z_\ast)$ is the comoving angular diameter distance and
$y(z) = \lambda_{21} (1+z)^2 / H(z)$.  $\Delta r_\parallel$ is the
comoving distance intervals corresponding to the frequency intervals
$\Delta f$. Then, the relation between ${\bm k}$ and ${\bm u}$ 
can be 
written as
\begin{equation}
{\bm u}_\perp = d_A {\bm k}_\perp, 
\qquad
u_\parallel = y k_\parallel.
\end{equation}
The power spectrum of $\delta T_b$ in $u$-space can be defined in
exactly the same manner as that for $k$-space by replacing ${\bm k}$
with ${\bm u}$ in Eq.~\eqref{eq:power}. By using the relation between
${\bm k}$ and ${\bm u}$, the power spectra in each space 
are connected
as
\begin{equation}
P_{21} ({\bm u})  = \frac{1}{d_A(z)^2  y (z)} P_{21} ({\bm k}).
\end{equation}
We use the $u$-space power spectrum in the following analysis.

With the power spectrum $ P_{21} ({\bm u}) $, the Fisher matrix is given by
\begin{equation}
\label{eq:Fisher_21}
F^{({\rm 21cm})}_{ij} = 
\sum_{\rm pixels}  
\frac{1}{[ \delta P_{21}({\bm u}) ]^2} 
\left( \frac{\partial P_{21} ({\bm u})}{\partial p_i} \right)
\left( \frac{\partial P_{21} ({\bm u})}{\partial p_j} \right),
\end{equation}
where $\delta P_{21}({\bm u})$ is the error in the
power spectrum measurements for a pixel ${\bm u}$, and $p_{i}$
represents cosmological parameters.  To be conservative, when we
differentiate $P_{21}({\bm u})$ with respect to cosmological
parameters, we fix ${\cal P}_{\delta \delta}(k)$ in
Eqs.~(\ref{eq:Pxx})~and~(\ref{eq:Pxdelta}) so that constraints only
come from the ${\cal P}_{\delta \delta}(k)$ terms in
$P_{\mu^0},P_{\mu^2},P_{\mu^4}$.  The error of the power spectrum
$\delta P_{21}({\bm u})$ comes from sample variance and experimental
noise, and is given by 
\begin{equation}
\delta P_{21}({\bm u}) = \frac{ P_{21}({\bm u})  + P_N (u_{\perp}) }{N_c^{1/2} }.
\end{equation}
The first term on the right hand side gives the one from the sample variance.
$N_c = 2 \pi k_\perp \Delta k_\perp \Delta k_\parallel V(z) /(2\pi)^3$
is the number of independent cells in an annulus summing over the
azimuthal angle.  Here $V(z) = d_A(z)^2 y(z) B \times {\rm FoV} $ is
the survey volume with $B$ being the bandwidth and 
FoV $\propto \lambda^{2}$ is the field of view of the interferometer.
The noise power spectrum, denoted as $P_N (u_{\perp})$ in the above formula, is given by
\begin{equation}
P_N (u_{\perp}) = \left( \frac{\lambda^{2} (z) T_{\rm sys} (z)  }{A_e (z)} \right)^2 
\frac{1}{t_0 n(u_\perp)},
\end{equation}
where $T_{\rm sys} = T_{{\rm sky}} + T_{{\rm rcvr}}$
($T_{{\rm sky}} = 60 (\lambda/[m])^{2.55} $ [K] : sky temperature, 
$T_{{\rm rcvr}} = 0.1T_{{\rm sky}} + 40 $[K] : receiver noise)
is the system temperature \cite{SKA},
which is dominated by the sky temperature due to synchrotron radiation, 
$A_e \propto \lambda^{2} $  is the effective collecting area,
$t_0$ is the  observation time and $n(u_\perp)$ is the number density of the
baseline, which depends on actual realization of antenna distributions
of each experiment.
\begin{table}[t]
  \centering 
  \begin{tabular}{cccccccc}
\hline \hline
Experiment & $N_{\rm ant}$& $A_e (z=8)$& $L_{\rm min}$
   & $L_{\rm max}$& ${\rm FoV}(z=8)$ & $t_{0}$&  $z$ \\ 
   & &$[{\rm m}^2]$&$[{\rm m}]$
   & $[{\rm km}]$ &$[{\rm deg}^2]$&[hour]&\\
\hline
SKA  phase1     & $911$ & $443$ & $35$ & $6$ & $13.12 \times 4 \times 4$  & 4000 & $6.8-10$ \\
Omniscope &  $10^6$ & $1$   & $1$    & $1$   & $2.063\times 10^4$ & 16000 & $6.8-10$  \\ 
\hline \hline
\end{tabular}
\caption{Specifications for interferometers of 21 cm experiments adopted in the analysis.
 For Omniscope, we assume the effective collecting area $A_{e}$ and  field of view are fixed.
  For SKA phase2,  it has the 10 times larger total collecting area than the phase1.
  Hence we take its noise power spectrum  to be 1/100 of the phase1, 
  and  the other specifications to be the same values. 
In addition, for SKA, we assume it uses 4 multi-beaming \cite{Mellema:2012ht},
and its total observation time is the same value as that of Omniscope (16000 hours),
but it observes 4 places in the sky (i.e. 4 times larger FOV and one fourth $t_{0}$.)
}
  \label{tab:21obs}
\end{table}

In our analysis, we consider the redshift range $ z = 6.75 - 10.05$,
which we divide into 4 bins: $z = 6.75 - 7.25, 7.25 - 7.75, 7.75-8.25$
and $8.25 -  10.05$.  For the wave number, we set $k_{{\rm
    min} \parallel} = 2 \pi / (y B) $ to avoid foreground
contamination \cite{McQuinn:2005hk} and take $ k_{\rm max} = 2~{\rm
  Mpc}^{-1}$ in order not to be affected by nonlinear effect which
becomes important on $k \ge k_{\rm max}$. 

To obtain the future cosmological constraints from 21 cm experiments,
we consider SKA (phase1, phase2) \cite{SKA,Mellema:2012ht} and Omniscope \cite{Tegmark:2009kv} 
whose specifications are shown in Table~\ref{tab:21obs}.
%
In order to calculate number density of baseline $n(u_\perp)$,
we assume a realization of antenna distributions for these arrays as follows.
For SKA phase1, we take 95\% (866) of the total antennae (stations)
distributed with a core region of radius 3000 m.
The distribution has an antenna density profile $\rho(r)$ ($r$: a radius from center of the array) as follows,
\begin{equation}
\rho(r) = 
\left\{
\begin{array}{lll}
 \rho_{0}r^{-1},     &\rho_{0} \equiv \frac{13}{16\pi\left(\sqrt{10}-1\right) }  \ {\rm m}^{-2}
& \hspace{60pt} r \leq 400 \ {\rm m},\\
 \rho_{1}r^{-3/2},  &\rho_{1} \equiv \rho_{0} \times 400^{1/2}, & \ \ \ 400 \ {\rm m} < r \leq 1000 \ {\rm m}, \\
 \rho_{2}r^{-7/2},  &\rho_{2} \equiv \rho_{1} \times 1000^{2}, & \ \ 1000 \ {\rm m} < r \leq 1500 \ {\rm m}, \\
 \rho_{3}r^{-9/2},  &\rho_{3} \equiv \rho_{2} \times 1500 ,          & \ \ 1500 \ {\rm m} < r \leq 2000 \ {\rm m}, \\
 \rho_{4}r^{-17/2},&\rho_{4} \equiv \rho_{3} \times 2000^{4}, & \ \ 2000  \ {\rm m} < r \leq 3000 \ {\rm m}. \\
\end{array}
\right.
\end{equation}
This distribution agrees with the specification of the SKA phase1 baseline design.
We ignore measurements from the sparse distribution of the remaining 5\% of the
total antennae that are outside this core region.
For SKA phase2, we assume that it has the 10 times larger total collecting area than the phase1.
Hence we take its noise power spectrum  to be 1/100 of the phase1.
We assume that the other specifications of SKA phase2 are the same as values of the phase1.
For Omniscope, which is a future square-kilometer collecting area array
optimized for 21 cm tomography, we take all of antennae distributed with a filled nucleus
 in the same manner as Ref. \cite{Mao:2008ug}.
In addition, we assume an azimuthally symmetric distribution of the antenna in both arrays.

\subsection{CMB}

As mentioned previously, although 21 cm experiments have strong power
to probe the primordial power spectrum, especially, on small scales,
observations of CMB greatly help to determine other cosmological
parameters such as energy densities of dark matter, baryon and dark
energy, and so on.  To obtain the future constraints, we consider
Planck \cite{Planck:2006aa}, CMBpol \cite{Baumann:2008aq}
and COrE \cite{Bouchet:2011ck}
whose specifications are summarized in Table~\ref{tab:cmb_obs}.

The Fisher information matrix for CMB is given by
\cite{Tegmark:1996bz,Zaldarriaga:1997ch,Oyama:2012tq}
\begin{equation}
F_{ij}^{\rm (CMB)}
= \sum_{l}
\frac{\left( 2l+1\right)}{2}f_{\mathrm{sky}}
\mathrm{Trace}
\left[
  \bm{\mathrm{C}}_{l}^{-1}
  \frac{\partial\bm{\mathrm{C}}_{l}}{\partial p_i}
  \bm{\mathrm{C}}_{l}^{-1}
  \frac{\partial \bm{\mathrm{C}}_{l}}{\partial p_j}
\right],
\end{equation}
where $\bm{\mathrm{C}}_{l}$ is a covariance matrix of CMB,
and $p_i$ represents cosmological parameters.
We take the maximum multipole to be $l_{\rm max} = 3000$ in the analysis. 
The covariance matrix $\bm{\mathrm{C}}_{l}$ is expressed as
\begin{equation}
\bm{\mathrm{C}}_{l} =  \left(
\begin{array}{ccc}
C_{l}^{TT} + N_{l}^{T} &
C_{l}^{TE} &
C_{l}^{Td} \\
\! \!  C_{l}^{TE}  &
C_{l}^{EE}+N_{l}^{P} &
0 \\
C_{l}^{Td}&
0 &
C_{l}^{dd}+N_{l}^{d}
\end{array}
\right),
\end{equation}
where $C^X_l \left(X=TT, EE, TE \right)$ is an angular power spectrum of each unlensed CMB mode,
$C_{l}^{dd}$ is one of deflection angle of CMB due to weak gravitational lensing,
$C_{l}^{Td}$ is cross correlation between temperature and deflection angle,
$N_{l}^{Y}\left(Y=T, P \right)$ 
is a noise power spectrum of temperature or polarization,
and $N_{l}^{d}$ is one of deflection angle.
The noise power spectrum $N^Y_l$ is given by,
\begin{equation}
N_l^{Y} (\nu)  = \Delta _Y^2 \exp \left[ l (l+1) \sigma_b^2 (\nu) \right],
\end{equation}
 for a
single frequency band,
where $\Delta_Y$ can be found in Table~\ref{tab:cmb_obs} and $\sigma_b
(\nu) = \theta_{\rm FWHM} / \sqrt{8 \ln 2}$ characterizes the width of
the beam.  When multiple frequency bands are used, 
where $N_l^Y$ is given by
the sum of its inverse as
\begin{equation}
\left( N_l^Y \right)^{-1}  = \sum_{\nu_i}  \frac{1}{N_l^Y (\nu_i)}.
\end{equation}
In addition, 
we estimate the noise power spectrum of deflection angle $N^{d}_l$
by assuming that we use lensing reconstruction with the quadratic estimator 
\cite{Okamoto:2003zw},
and compute it by using a public code FUTURCMB \cite{paper:FUTURCMB},
which adopts the estimator.

Equipped with these formalism, we can calculate the Fisher matrix to
obtain projected constraints on the spectral index $n_s$ and the
runnings $\alpha_s$ and $\beta_s$, which will be presented in the next
subsection.

\begin{table}[tbp]
  \centering
  \begin{tabular}{c|cccc}
  \hline \hline
experiment &  frequency  & beam  $\theta_{\rm FWHM}$ & $\Delta_T$ & $\Delta_P$ \\
           &[GHz] & [arcmin]
           &[$\mu$K arcmin] &  [$\mu$K arcmin] \\
\hline
Planck
& 100 & 9.5 & 64.6 & 104 \\
& 143 & 7.1 & 42.6 &  80.9\\ 
& 217 & 5 & 65.5  & 134\\ 
\hline
CMBpol 
& 45  & 17 & 5.85   & 8.27 \\
& 70  & 11 & 2.96   & 4.19 \\
& 100 & 8   & 2.29   & 3.24 \\
& 150 & 5   & 2.21   & 3.13 \\
& 220 & 3.5& 3.39   & 4.79 \\
\hline
COrE 
& 75  & 14    & 2.73   & 4.72 \\
& 105 & 10   & 2.68   & 4.63 \\
& 135 & 7.8  & 2.63   & 4.55 \\
& 165 & 6.4  & 2.67   & 4.61 \\
& 195 & 5.4  & 2.63   & 4.54 \\
& 225 & 4.7  & 2.64   & 4.57 \\
\hline \hline
\end{tabular} 
\caption{
Specifications for Planck, CMBpol and COrE adopted in the analysis. 
For CMBpol, we assumed the mid-cost (EPIC-2m)  mission and 
only used five frequency bands 
for a realistic foreground removal.
For the same reason, for COrE, we only used six frequency bands.
In the both observations,  $f_{\rm sky} =0.65$ is assumed.
}
  \label{tab:cmb_obs}
\end{table}

\subsection{Future constraints}

Now we present our results for projected constraints on cosmological
parameters, paying particular attention to $n_s, \alpha_s$ and
$\beta_s$ which characterize the scale-dependence of primordial
curvature perturbations.  In addition to these parameters, we also
vary the standard cosmological parameters.  Thus we explorer an
8-dimensional parameter space: $  \Omega_{\Lambda},
\Omega_bh^2, h, \tau, A_s, n_s, \alpha_s, \beta_s$, 
where  $\Omega_{\Lambda}$ and $\Omega_b$ are
energy densities of cosmological constant
and baryon, respectively, $h$ is the
Hubble parameter in units of $ 100 {\rm km}/{\rm s}/{\rm Mpc}$ and
$\tau$ is the reionization optical depth. In the analysis, we assume a
flat universe, and fix the Helium abundance to be $Y_p =
0.25$. Neutrinos are treated as massless particles and its effective
number is also fixed to be $N_{\rm eff} = 3.046$.
$A_s, n_s, \alpha_s$ and $\beta_s$ are parameters which characterize the primordial power
spectrum as already mentioned,
and we normalize its  amplitude 
as  $A_{s}(k_{\rm ref}) \equiv {\cal P}_{\zeta}(k_{\rm ref})/(2.21381 \times 10^{-9} )$ 
\cite{Verde:2003ey,Peiris:2003ff}.
As a reference scale to parametrize
these quantities, we take $k_{\rm ref} = 0.05~{\rm Mpc}^{-1}$ in the
most analysis presented below.  However, we also discuss how this
reference scale affects the determinations of $n_s, \alpha_s$ and
$\beta_s$ at the final part of this section.
Then we set the fiducial values of these parameters (except $\alpha_{s}, \beta_{s}$) 
near the best fit of the  Planck results (Planck + WMAP polarizations + high L CMB data + BAO)  \cite{Ade:2013zuv} ,
and $\alpha_{s}=0, \beta_{s}=0$, 
so that $( \Omega_{\Lambda},\Omega_bh^2, h, \tau, A_s, n_s, \alpha_s, \beta_s)$
$=(0.6914, 0.022161, 0.6777, 0.0952, 1, 0.9611, 0,0)$.

\begin{figure}[t]
  \begin{center}
    \resizebox{120mm}{!}{
    \includegraphics{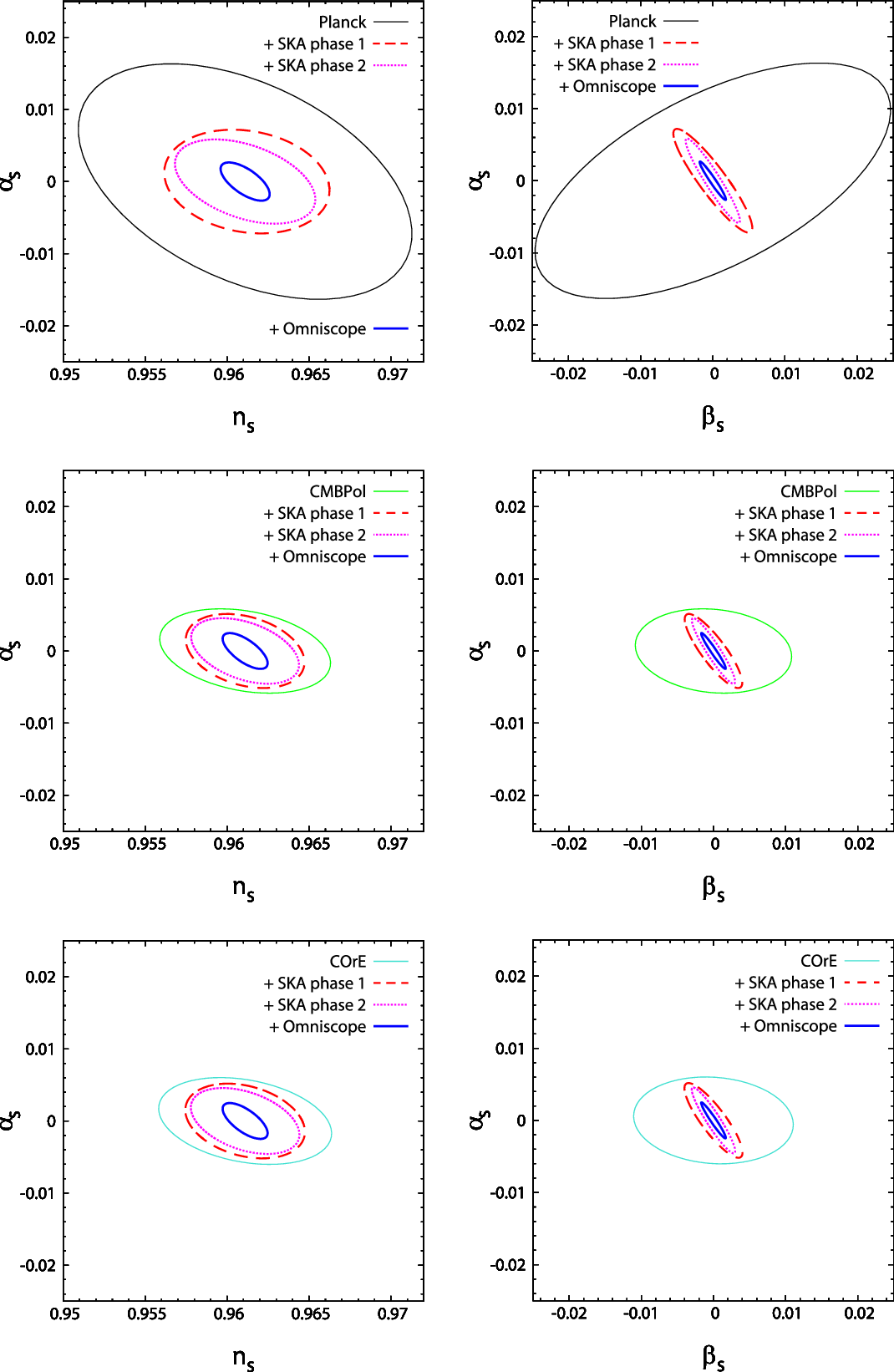}
  }
  \end{center}
  \caption{
Projected constraints from Planck, Planck+SKA phase1, Planck+SKA phase2,
 Planck+Omniscope (top panels) , CMBpol, CMBpol+SKA phase1, CMBpol+SKA phase2,
 CMBpol+Omniscope (middle panels) and COrE, COrE+SKA phase1, COrE+SKA phase2,
 COrE+Omniscope (bottom panels) in the $n_s$--$\alpha_s$ (left)
    and $\alpha_s$--$\beta_s$ (right) planes. 
}
  \label{fig:comparison}
\end{figure}

The total Fisher matrix is given by 
\begin{equation}
F_{ij} = F^{\rm (21cm)}_{ij} + F^{\rm (CMB)}_{ij},
\end{equation}
where $i,j$ are subscript representing cosmological parameters.  We
report our results for several combinations of experiments. As already
mentioned above, although 21 cm experiments enable us to probe
the primordial power spectrum very precisely since
they can measure fluctuations smaller scales compared to those of CMB,
regarding the determinations of other cosmological parameters, CMB has
more strong power. Since the scale-dependence of the
power spectrum has degeneracies
with other cosmological parameters, reducing the uncertainties of such parameters
is also  important to precisely measure $n_s, \alpha_s$ and $\beta_s$.

\begin{table}[tbp]
  \centering
  \begin{tabular}{l|ccc}
  \hline \hline \\
   &  $ \delta n_s $ &  $ \delta \alpha_s $ & $ \delta \beta_s $ \\ \hline
 Planck 
 & $4.11 \times 10^{-3}$& $6.59 \times 10^{-3}$ & $9.95 \times 10^{-3}$   \\ 
Planck + SKA phase1
& $2.03 \times 10^{-3}$  & $2.90 \times 10^{-3}$   &  $2.21 \times 10^{-3}$   \\
Planck + SKA phase2
& $1.73 \times 10^{-3}$  & $2.36 \times 10^{-3}$   &  $1.52 \times 10^{-3}$   \\
Planck + Omniscope 
& $6.04 \times 10^{-4}$  &  $1.07 \times 10^{-3}$  &  $7.31 \times 10^{-4}$   \\ \hline
CMBpol
 &  $2.10 \times 10^{-3}$  &  $2.36 \times 10^{-3}$  &  $4.37 \times 10^{-3}$  \\ 
CMBpol + SKA phase1
& $1.46 \times 10^{-3}$  & $2.07 \times 10^{-3}$   &  $1.61 \times 10^{-3}$   \\
CMBPol + SKA phase2
& $1.33 \times 10^{-3}$  & $1.84 \times 10^{-3}$   &1.21  $ \times 10^{-3}$   \\
CMBpol + Omniscope 
& $5.53 \times 10^{-4}$  &  $1.00 \times 10^{-3}$  &  $6.86 \times 10^{-4}$   \\ \hline 
COrE
 &  $2.13 \times 10^{-3}$  &  $2.43 \times 10^{-3}$  &  $4.47 \times 10^{-3}$  \\ 
COrE + SKA phase1
& $1.47 \times 10^{-3}$  & $2.09 \times 10^{-3}$   &  $1.63 \times 10^{-3}$   \\
COrE + SKA phase2
& $1.34 \times 10^{-3}$  & $1.85 \times 10^{-3}$   &  $1.22 \times 10^{-3}$   \\
COrE + Omniscope 
& $5.54 \times 10^{-4}$  &  $1.00 \times 10^{-3}$  &  $6.87 \times 10^{-4}$   \\ \hline \hline 
\end{tabular} 
\caption{1$\sigma$ uncertainties for $n_s, \alpha_s$ and $\beta_s$ from various data sets.
  We take $k_{\rm ref} = 0.05~{\rm Mpc}^{-1}$ to derive these constraints.}
  \label{tab:summary}
\end{table}

In Fig.~\ref{fig:comparison}, we show projected constraints in the
$n_s$--$\alpha_s$ and $\alpha_s$--$\beta_s$ planes where 2$\sigma$
limits are shown for the analysis from 
Planck, Planck + SKA (phase1 or phase2), Planck + Omniscope, 
CMBpol, CMBpol + SKA (phase1 or phase2), CMBpol + Omniscope 
and COrE, COrE + SKA (phase1 or phase2), COrE + Omniscope.
Uncertainties for each parameter $n_s, \alpha_s$ and $\beta_s$ are summarized in
Table~\ref{tab:summary}. We note that the 1$\sigma$ uncertainties are reported  in the table.
Although CMB experiments can already give a
very precise measurement of $n_s$ at better than $\mathcal{O} (1 \%)$,
when one includes the data from 21 cm fluctuations, especially with
Omniscope, the precision improves by even one order of magnitude as
seen from the figure and the table.  The same also holds for the
runnings $\alpha_s$ and $\beta_s$.  In particular, for the case with
CMBpol or COrE + Omniscope, one can probe the runnings $\alpha_s$ and $\beta_s$
with the precision down to $\mathcal{O}(10^{-4})$. As discussed in the
previous section, some inflationary model predicts $\alpha_s =
\mathcal{O}(10^{-3})$ and $\beta_s = \mathcal{O}(10^{-4})$, thus such models 
can be tested with future experiments of CMB and 21 cm fluctuations by using 
the runnings of the power spectrum.

Our results show that the scale-dependence of the primordial power
spectrum can be well probed when one uses the data from 21 cm
fluctuations, which is consistent with
\cite{Mao:2008ug,Barger:2008ii}, and further demonstrate that the
information of ``higher order" scale-dependence such as the running
$\alpha_s$ and the running of the running $\beta_s$ can also give
significant information on models of primordial density fluctuations.

\begin{figure}[htbp]
\begin{center}
\resizebox{120mm}{!}{ \includegraphics{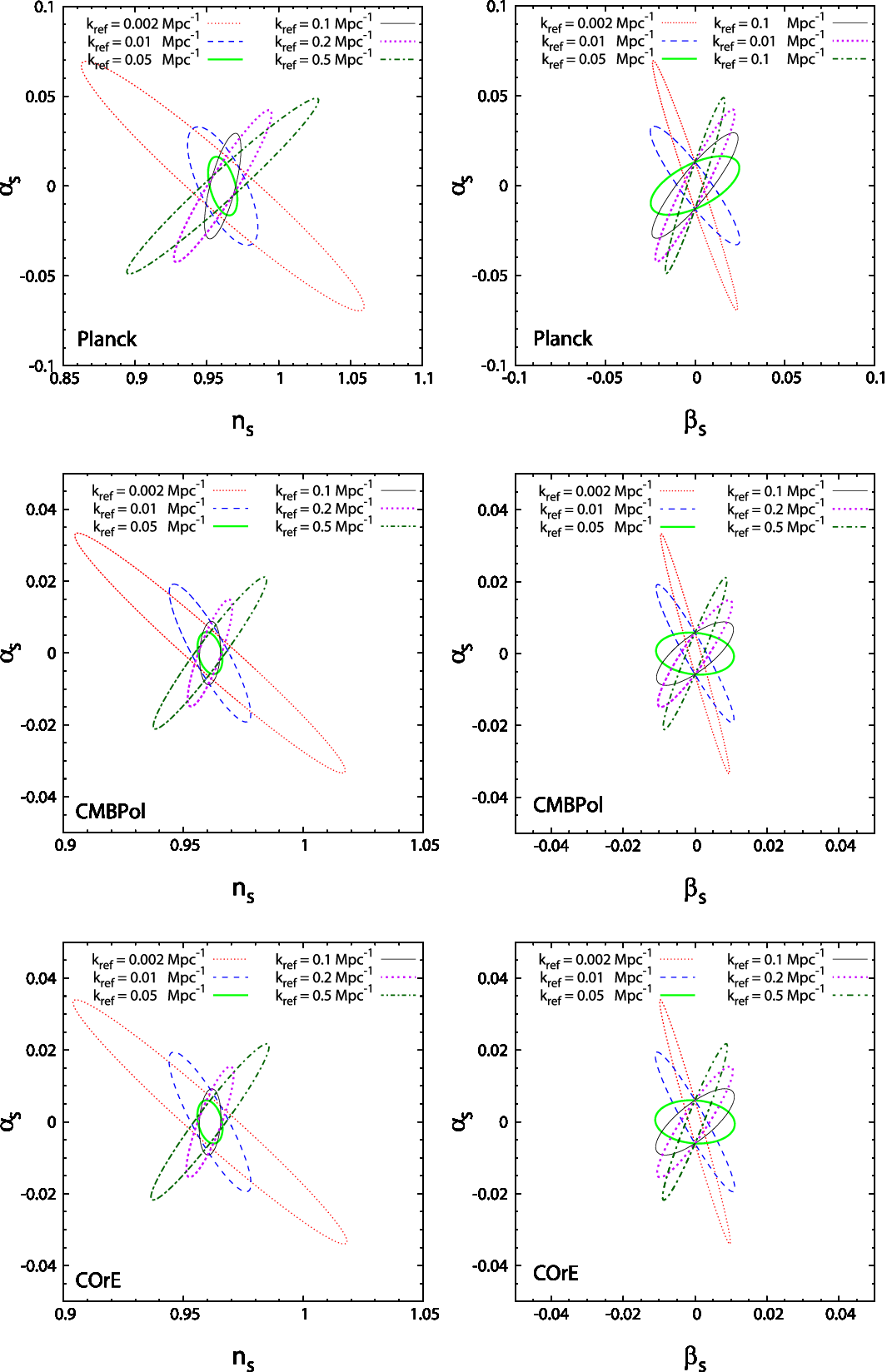} }
\caption{
Projected constraints in the $n_s$--$\alpha_s$ (left panels)
and $\alpha_s$--$\beta_s$ (right panels) planes for several
values of $k_{\rm ref}$ from Planck(top), CMBpol(middle) and  COrE(bottom). 
}\label{fig:CMB_k_ref}
\end{center}
\end{figure}

\begin{figure}[htbp]
\begin{center}
\resizebox{120mm}{!}{\includegraphics{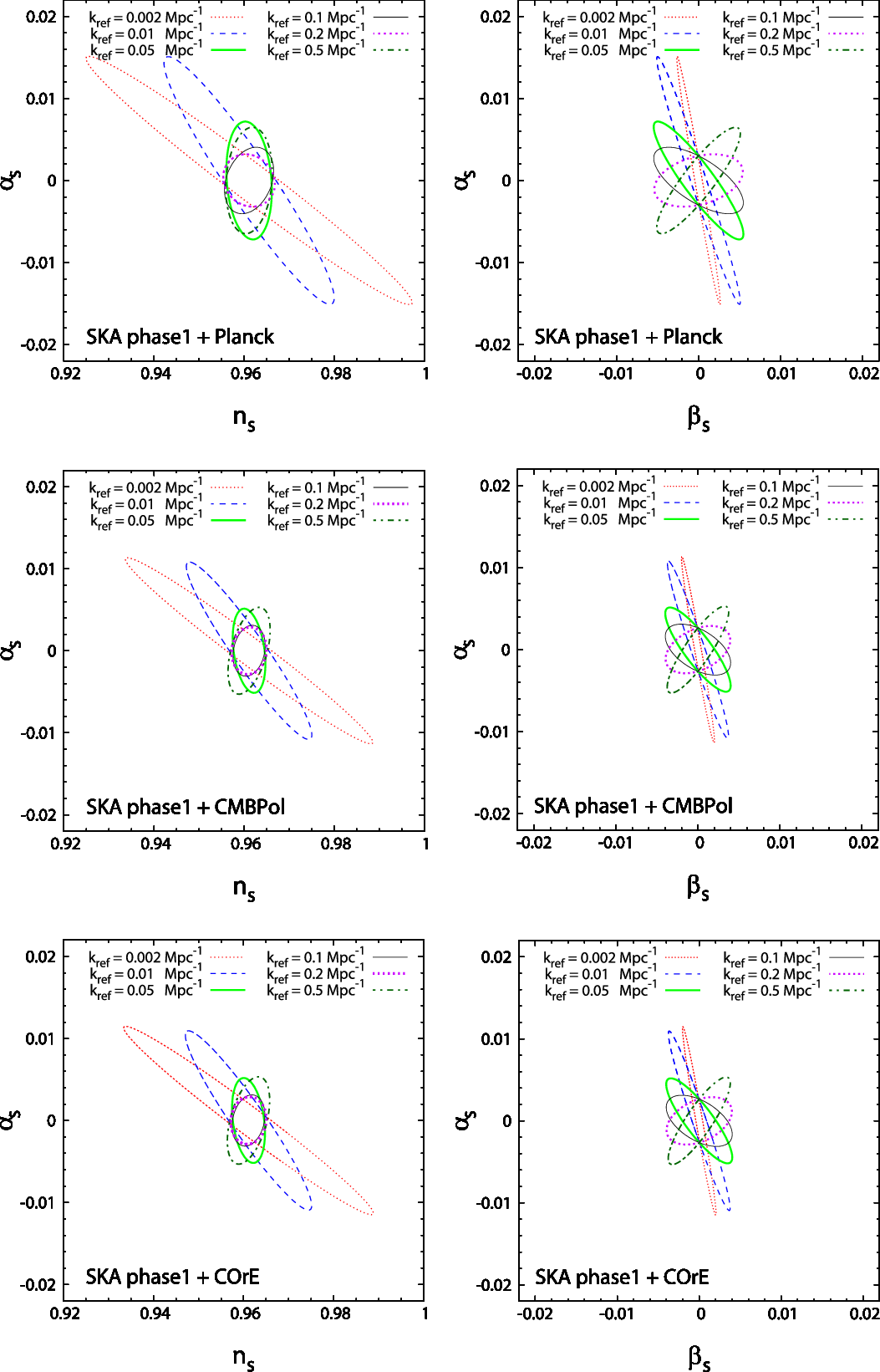}}
\caption{Same as Fig.~\ref{fig:CMB_k_ref}, but from Planck+SKA phase1(top), 
CMBpol+SKA phase1(middle) and COrE+SKA phase1(bottom).
}\label{fig:SKA1_k_ref}
\end{center} 
\end{figure}

\begin{figure}[htbp]
 \begin{center}
\resizebox{120mm}{!}{\includegraphics{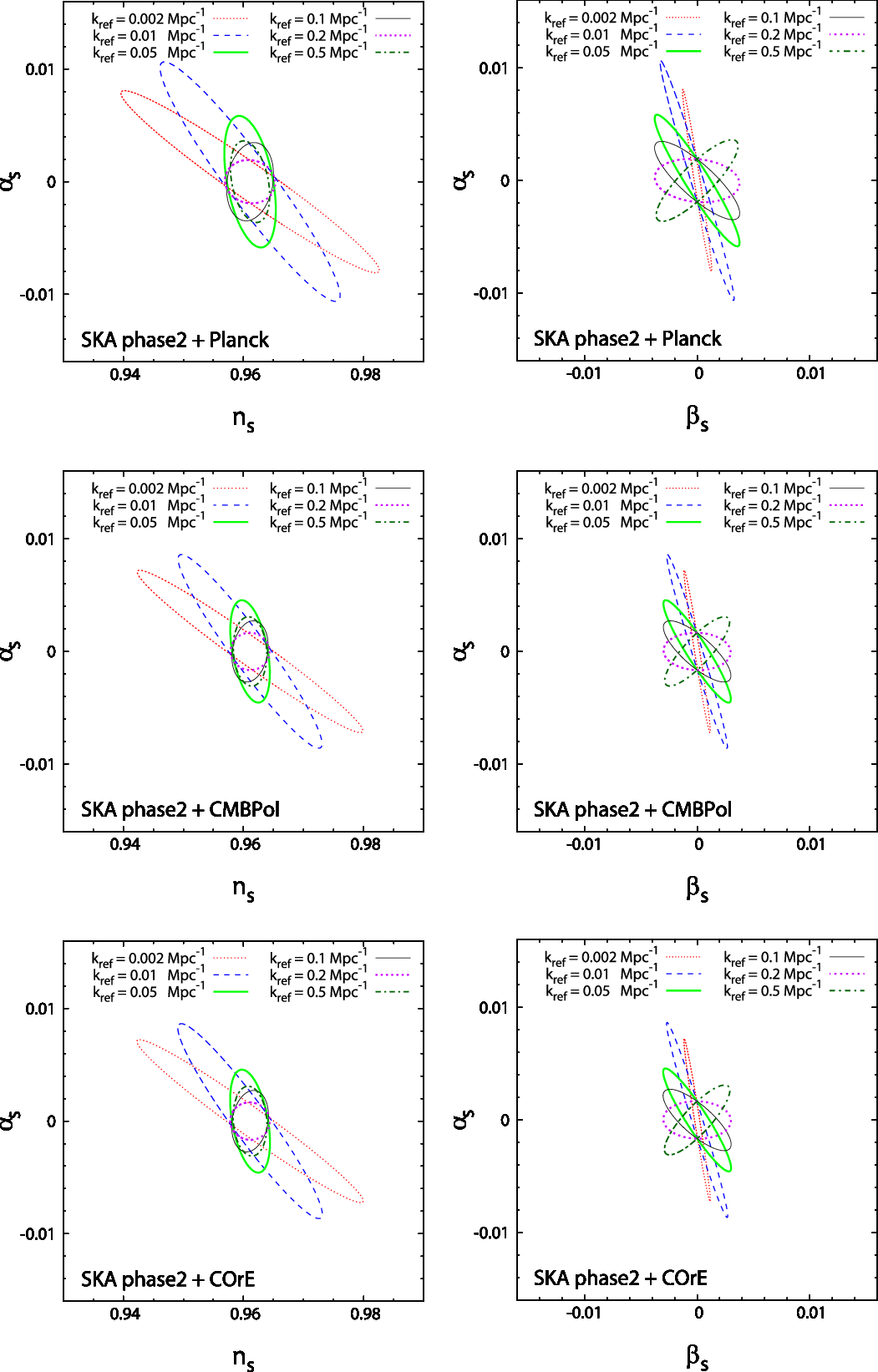}}
\caption{Same as Fig.~\ref{fig:CMB_k_ref}, but from Planck+SKA phase2(top), CMBpol+SKA phase2(middle) and 
COrE+SKA phase2(bottom).
 }\label{fig:SKA2_k_ref}
\end{center}
\end{figure}

\begin{figure}[htbp]
\begin{center}
\resizebox{120mm}{!}{\includegraphics{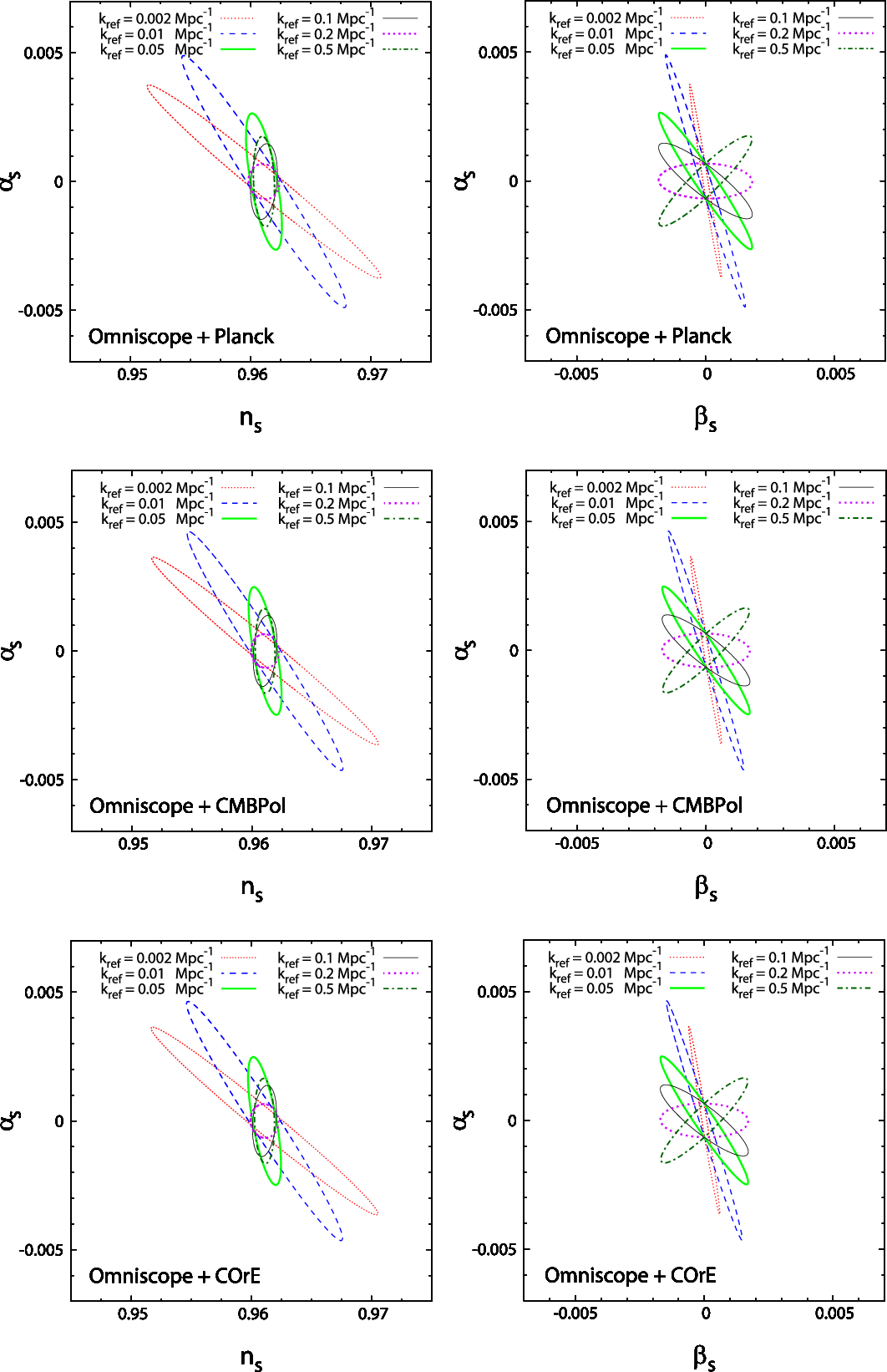}}
\caption{Same as Fig.~\ref{fig:CMB_k_ref}, but 
from Planck+Omniscope (top),
CMBpol+Omniscope (middle) and COrE+Omniscope(bottom). 
}\label{fig:kref_Omni}
\end{center}
\end{figure}

Finally we discuss how the choice of $k_{\rm ref}$ affects the
determination of the parameters $n_s, \alpha_s$ and $\beta_s$. In
Fig.~\ref{fig:kref_Omni}, expected constraints on the
$n_s$--$\alpha_s$ and $\alpha_s$--$\beta_s$ planes are shown for
several values of $k_{\rm ref}$ for 
Planck + Omniscope, CMBpol + Omniscope, COrE + Omniscope.
1$\sigma$  uncertainties for each parameter are
summarized in Table~\ref{tab:summary2}.  As seen from the figure, the
choice of $k_{\rm ref}$ affects the uncertainties for $n_s, \alpha_s$
and $\beta_s$ by a factor of a few and also changes the direction of
the degeneracy. From the viewpoint of determining the scale dependence
parameters, the optimal reference scale would be the one which gives
uncorrelated constraints among the parameters. However, by looking at
Fig.~\ref{fig:kref_Omni}, we can see that the reference scale giving
uncorrelated measures for the parameter sets $(n_s, \alpha_s)$ and
$(\alpha_s, \beta_s)$ are different. In addition, even just
considering CMB or 21 cm experiment alone, the optimal scale seems to
also depend on the specification of the experiments.  But a general
tendency is that the optimal scales is around from $0.05~{\rm
  Mpc}^{-1}$ to $0.1~{\rm Mpc}^{-1}$. Thus we have mainly used the
reference scale $k_{\rm ref} =0.05~{\rm Mpc}^{-1}$ to show our
results.

\begin{table}[tbp]
  \centering
  \begin{tabular}{c|cccccc}
  \hline \hline \\
 $k_{\rm ref} $ 
&  $0.002 $ &  $0.01 $  & $0.05 $ & $0.1 $& $0.2 $& $0.5 $ \\ 
$[{\rm Mpc}^{-1}]$& \\ \hline
$ \delta n_s $
& $3.81 \times 10^{-3}$ & $2.62 \times 10^{-3}$ &  $5.53 \times 10^{-4}$  & $4.01 \times 10^{-4}$  & $4.68 \times 10^{-4}$  & $3.33 \times 10^{-4}$\\ 
$ \delta \alpha_s $
& $1.47 \times 10^{-3}$ & $1.87 \times 10^{-3}$  &  $1.00 \times 10^{-3}$  &  $5.57 \times 10^{-4}$  & $2.64 \times 10^{-4}$  & $6.65 \times 10^{-4}$  \\
$ \delta \beta_s $ 
& $2.43 \times 10^{-4}$  & $5.94 \times 10^{-4}$ &  $6.86 \times 10^{-4}$  &  $6.88 \times 10^{-4}$  & $6.87 \times 10^{-4}$  & $6.79 \times 10^{-4}$ 
\\ \hline \hline 
\end{tabular} 
\caption{Expected 1$\sigma$ uncertainties of $n_s, \alpha_s$ and $\beta_s$
  from CMBpol+Omniscope  for several values of $k_{\rm ref}$.}
  \label{tab:summary2}
\end{table}

\section{Conclusion}

We have investigated how precisely one can measure the power spectrum
of the curvature perturbation from future experiments of 21 cm
fluctuations and CMB.  In particular, we have studied projected
constraints on the parameters characterizing the scale-dependence of
the power spectrum such as the spectral index $n_s$, its running
$\alpha_s$ and the running of the running $\beta_s$.  Although the
former two parameters have been well explored in various context in
the literature, the latter one, $\beta_s$, has not been studied much
in connection with  cosmological probes.

Although the gravity waves or the tensor mode can give significant
information to the inflationary Universe once it is detected, there
are many inflation models, such as small-field models, predicting too
small tensor-to-scalar ratio. In addition, in models with a light
scalar field such as the curvaton model, modulated reheating scenario
and so on, which are of interest due to the possibilities of their
giving large $f_{\rm NL}$, the tensor-to-scalar ratio also tends to be
very small. If one of these models is realized in the nature, it would
be very difficult to detect the signature from the gravity
waves. However, even in that case, ``higher order" scale dependence of
(scalar) curvature perturbations would help to probe the inflationary
model, which we show quantitatively  in this paper. We have discussed some explicit
models where higher order running $\beta_s$ would be very useful to
differentiate models.  Needless to say, even when the tensor modes are
detected, the runnings can give extra valuable information on models
of primordial fluctuations.

We have  obtained expected constraints on such a
higher order running $\beta_s$ as well as $n_s$ and $\alpha_s$ by
using observations of 21 cm fluctuation, in combination with CMB.  Since
the power spectrum of 21 cm fluctuations can probe cosmic density
fluctuations on smaller scales than those observed in CMB,  
one can obtain severe constraints even for the
running of running parameter $\beta_s$.  In particular, when one
considers the combination of 
CMBpol or COrE + Omniscope, 
we can probe the running parameters with the precision of $\delta \alpha_s =
\mathcal{O}(10^{-3})$ and $\delta \beta_s = \mathcal{O}(10^{-4})$,
which would give useful information to discriminate inflationary
models.

Although current cosmological observations are already so precise that
some of inflationary models have been excluded, there are still many
possibilities allowed and thus we need to go further to understand the
early Universe more.  In particular, to see the details and
differentiate models well, it is preferable to have yet another
observables other than commonly used one. As such a quantity, we
considered a higher order running $\beta_s$ in this paper, which can
be well probed by future cosmological observations such as from 21 cm
and CMB.  In the view that cosmological data will be much more precise
in the future, the research along this line would provide us a lot of
insight on the inflationary Universe and even the origin of the
Universe.

\section*{Acknowledgments}
We thank Garrelt Mellema for a useful correspondence.
This work is partially supported by the Grant-in-Aid for Scientific
research from the Ministry of Education, Science, Sports, and Culture,
Japan, Nos. 21111006, 22244030, 23540327 (K.K.), 23.5622 (T.S.) and
23740195(T.T.).


\end{document}